\documentclass{article}
\usepackage[utf8]{inputenc}
\usepackage{amsmath}
\usepackage[margin=1in]{geometry}
\usepackage{graphicx}

\usepackage{color}
\usepackage{ulem}

\title{Warning of a forthcoming collapse of the Atlantic meridional overturning circulation}
\author{P. Ditlevsen$^1$ and S. Ditlevsen$^2$\\1. Niels Bohr Institute, University of Copenhagen\\2. Institute of Mathematical Sciences, University of Copenhagen}
\date{}

\begin{document}

\maketitle
\begin{abstract}
Tipping to an undesired state in the climate when a control parameter slowly approaches a critical value is a growing concern with increasing greenhouse gas concentrations. Predictions rely on detecting early warning signals (EWSs) in observations of the system. The primary EWSs are increase in variance, (loss of resilience), and increased autocorrelation (critical slow down). These measures are statistical in nature, which implies that the reliability and statistical significance of the detection depends on the sample size in observations and the magnitude of the change away from the base value prior to the approach to the tipping point. Thus, the possibility of providing useful early warning depends on the relative magnitude of several interdependent time scales in the system. These are (a) the time before the critical value is reached, (b) the (inverse) rate of approach to the tipping point, (c) the size of the time window required to detect a significant change in the EWS and finally, (d) the escape time for noise-induced transition (prior to the tipping). Conditions for early warning of tipping of the Atlantic meridional overturning circulation (AMOC) are marginally fulfilled for the existing past $\sim$150 years of proxy observations where indicators of tipping have recently been reported. Here we provide statistical significance and data driven estimators for the time of tipping. We estimate a collapse of the AMOC to occur around the year 2057 under the assumption of a "business as usual" scenario of future emissions.           
\end{abstract}

\subsection*{}
A forthcoming collapse of the Atlantic meridional overturning circulation (AMOC) is a major concern as it is one of the most important tipping elements in Earth's climate system~\cite{manabe:1988,rahmstorf:1995,lenton:2008}. In recent years, model studies and paleoclimatic reconstructions indicate that the strongest abrupt climate fluctuations, the Dansgaard-Oeschger events~\cite{dansgaard:1993}, are connected to the bimodal nature of the AMOC~\cite{vettoretti:2022,ganopolski:2001}. Numerous climate model studies show a hysteresis behaviour, where changing a control parameter, typically the freshwater input into the Northern Atlantic, makes the AMOC bifurcate through a set of co-dimension one saddle-node bifurcations~\cite{woods:2019,hawkins:2011,weijer:2019}. State-of-the-art Earth-system models can reproduce such a scenario, but inter-model spread is large and the critical threshold is poorly constrained~\cite{mecking:2017,rahmstorf:2015,ipcc:2021}.

When complex systems undergo critical transitions by changing a control parameter $\lambda$ through a critical value $\lambda_c$, a structural change in the dynamics happens. The previously statistically stable state ceases to exist and the system moves to a different statistically stable state. The system undergoes a bifurcation, which for $\lambda$ sufficiently close to $\lambda_c$ can happen in a limited number of ways rather independent from the details in the governing dynamics~\cite{guckenheimer:1986}.
Beside a decline of the AMOC before the critical transition, there are EWSs, statistical quantities, which also change before the tipping happens. These are critical slow down (increased auto-correlation) and, from the Fluctuation-Dissipation Theorem, increased variance in the signal~\cite{kubo:1966,ditlevsen:2010,boulton:2014}. The latter is also termed "loss of resilience", especially in the context of ecological collapse~\cite{scheffer:2009}. The two EWSs are statistical equilibrium concepts. Thus, using them as actual predictors of a forthcoming transition, rely on the assumption of quasi-stationary dynamics. 

\begin{center}
\begin{figure}[htbp]
\begin{center}
\includegraphics[width=12cm]{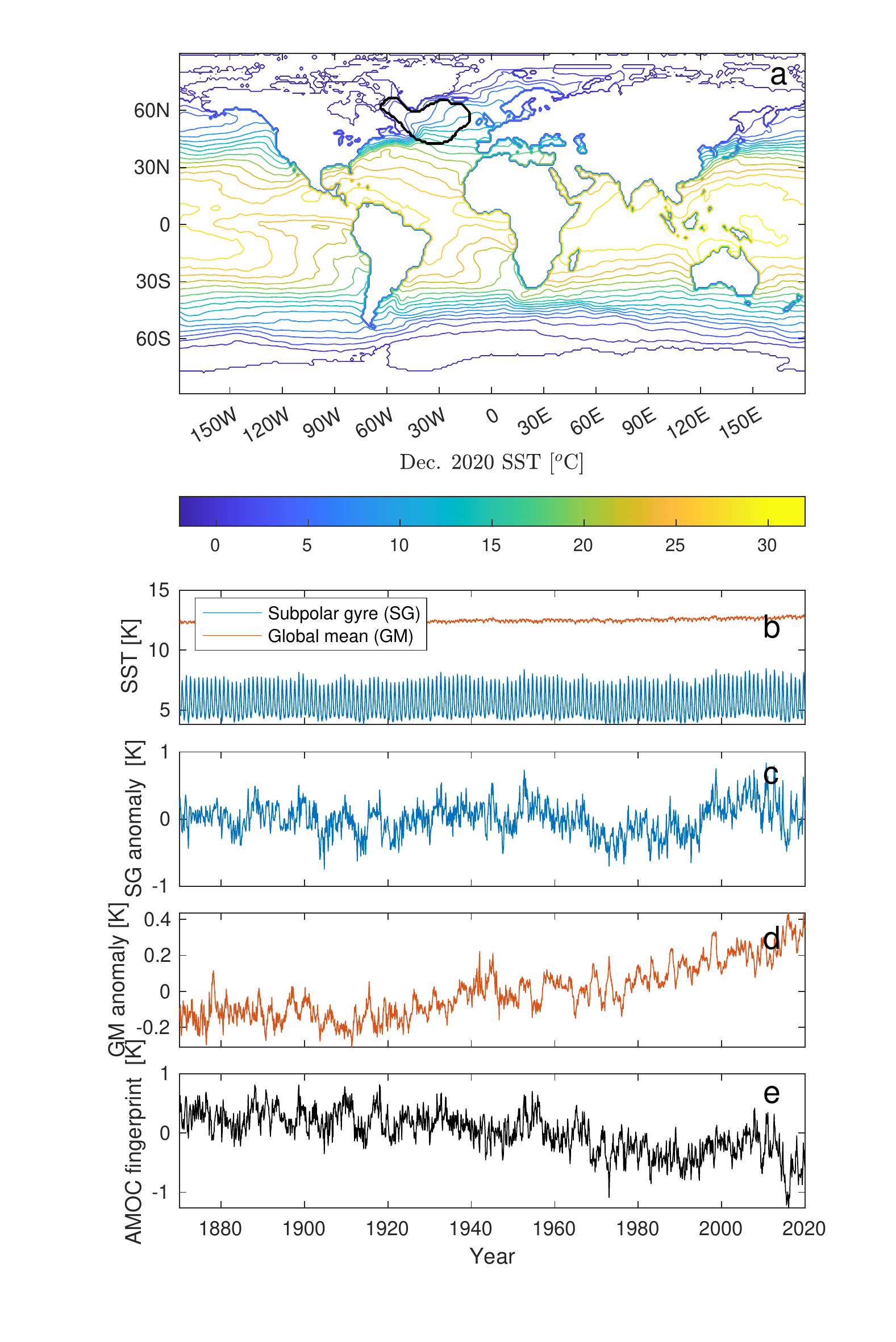}
\end{center}
\caption{\label{data} Panel {\bf a} shows the Subpolar gyre (SG) region (black contour) on top of the HasISST SST reconstruction for Dec. 2020. The SG region SST has been identified as an AMOC fingerprint \cite{caesar:2018}. Panel {\bf b} shows full monthly record of the SG SST together with the global mean (GM) SST. Panels {\bf c} and {\bf d} show the SG and GM anomalies, which are the records subtracted the monthly mean over the full record. Panel {\bf e} shows the AMOC fingerprint proxy, which is here defined as the SG anomaly minus twice the GM anomaly, compensating for the polar amplified global warming.}
\end{figure}
\end{center}

The AMOC has only been monitored continuously since 2004 through combined measurements from moored instruments, induced electrical currents in submarine cables and satellite surface measurements~\cite{smeed:2014}. Over the period 2004-2012 a decline in the AMOC has been observed, but longer records are necessary to assess the significance. For that, careful fingerprinting techniques have been applied to longer records of sea surface temperature (SST), which, backed by a survey of a large ensemble of climate model simulations, have found the SST in the Subpolar gyre (SG) region of the North Atlantic (Area marked with a black contour in Fig. \ref{data}{\bf a}) to contain an optimal fingerprint of the strength of the AMOC~\cite{caesar:2018,jackson:2020,latif:2004}. To obtain the AMOC fingerprint, two steps are required: The seasonal cycle in the SST is governed by the surface radiation independent from the circulation and thus removed by considering the monthly anomalies, where the mean over the period of recording of the month is removed. Secondly, there is an ongoing positive linear trend in the SST related to global warming, which is also not related to the circulation. This is compensated for by subtracting $2 \times$ the global mean (GM) SST anomaly (small seasonal cycle removed). This differs slightly from ref. \cite{rahmstorf:2015}, where $1\times$ the GM SST was subtracted. The factor 2 is the optimal value for the polar amplification~\cite{holland:2003} obtained by calibrating to recent direct measurements \cite{frajka-williams:2019} (supplementary text S6).

Fig. \ref{data}{\bf b} shows the SG and the GM SSTs obtained from the Hadley Centre Sea Ice and Sea Surface Temperature data set (HadISST)~\cite{rayner:2003}. Fig. \ref{data}{\bf c} shows the SG anomaly and Fig. \ref{data}{\bf d} shows the GM anomaly with a clear global warming trend in the last half of the record. The AMOC fingerprint for the period 1870-2020 is shown in Fig. \ref{data}{\bf e}. This is the basis for the analysis. It has been reported~\cite{rahmstorf:2015,boers:2021} that this and similar AMOC indices show significant trends in the mean, the variance and the autocorrelation, indicating early-warning of a shutdown of the AMOC. However, a trend in the EWSs within a limited period of observation could be a random fluctuation within a steady state statistics. Thus, for a robust assessment of the shutdown, it is necessary to establish a statistical confidence level for the change above the natural fluctuations. This is not easily done given only one, the observed, realization of the approach to the transition. Here we establish such a measure of the confidence for the variance and autocorrelation and demonstrate that variance is the more reliable of the two. A further contribution is an estimator of not only whether a transition is approaching, but also the time when the critical transition is expected to occur. We find that the transition where a control parameter reaches the critical value, is most likely to occur around 2057 with 95\% confidence interval 2034-2128. The strategy is to infer the evolution of the AMOC solely on observed changes in mean, variance and autocorrelation. The typical choice of control parameter is the flux of freshwater into the North Atlantic. River runoff, Greenland ice melt and export from the Arctic ocean are not well constrained~\cite{yang:2016}, thus we do not assume the control parameter known. Boers~\cite{boers:2021} assumes the global mean temperature $T$ to represent the control parameter. $T$ increases roughly linear with time since $\sim 1920$ (Fig. \ref{data}{\bf d}). All we assume here is that the AMOC is in an equilibrium state prior to a change towards the transition. The simplest uninformed assumption is that the change is sufficiently slow and that the control parameter approach the (unknown) critical value linearly with time. This assumption is confirmed a posteriory by a close fit to the observed AMOC fingerprint.        
\subsection*{Modeling and detecting the critical transition}
\begin{center}
\begin{figure}[htbp]
\begin{center}
\includegraphics[width=12cm]{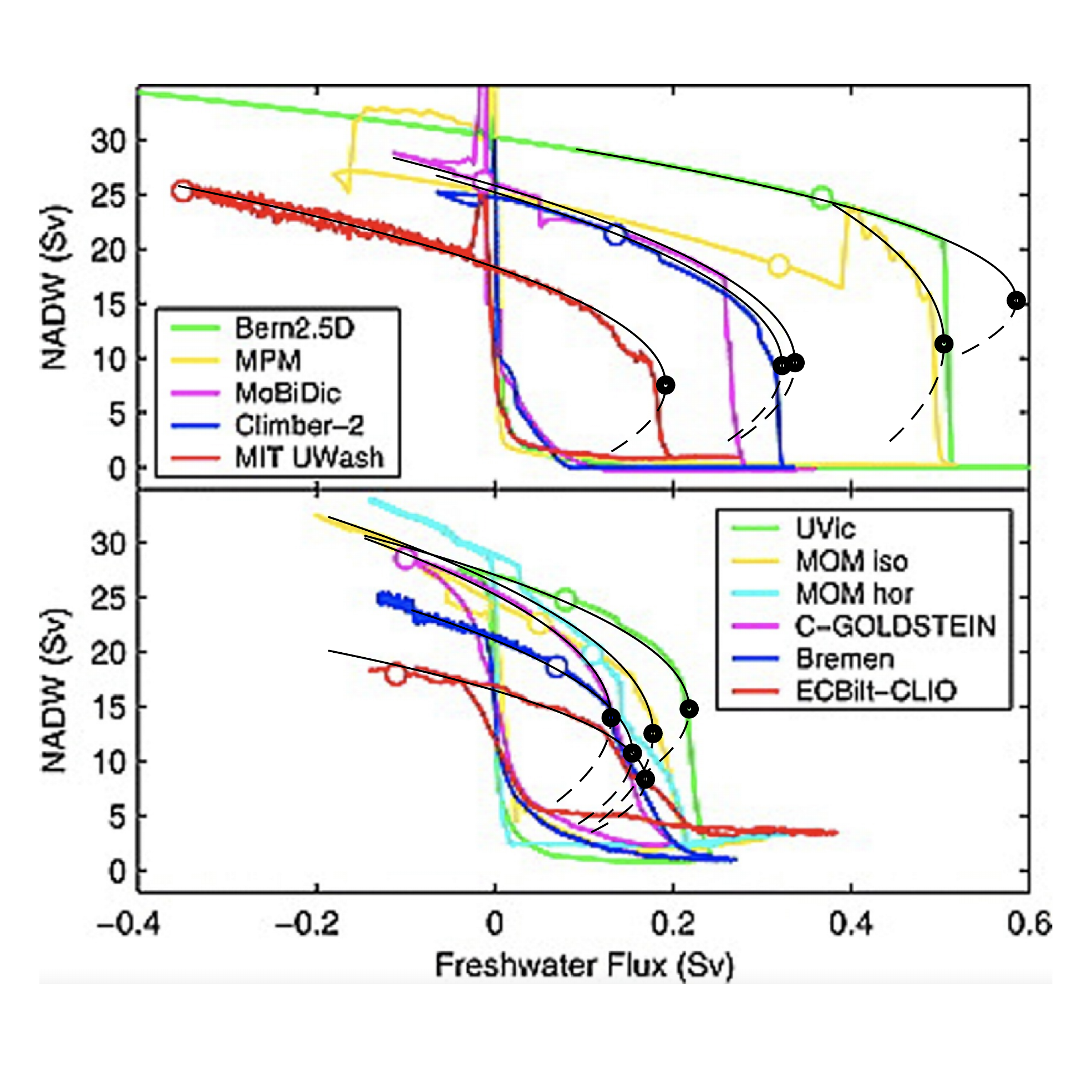}
\end{center}
\caption{\label{fig:sn2} 
The steady state curves from climate model simulations of the North Atlantic Deep Water (NADW), with a very slowly changing control parameter (freshwater forcing). Top panel shows ocean only models, while bottom panel shows atmosphere-ocean models. The curves are, even away from the transition surprisingly well fitted by eq. \eqref{eq:X} (black thin curves). The bifurcation points are indicated with black circles. Note that for some models the transition happens before the critical point, as should be expected from noise induced transitions. The colored circles show the present day conditions for the different models. Adapted from Rahmstorf et al. \cite{rahmstorf:2005}.}
\end{figure}
\end{center}

Denote the observed AMOC fingerprint by $x(t)$ (Fig. \ref{data}{\bf e}). We model it by a stochastic process $X_t$ which, depending on a control parameter $\lambda<0$, is in risk of undergoing a critical transition through a saddle-node bifurcation for $\lambda=\lambda_c=0$. The system is initially in a statistically stable state, i.e., it follows some stationary distribution with constant $\lambda = \lambda_0$.  We are uninformed about the dynamics governing the evolution of $X_t$, but can assume an effective dynamics, which, with $\lambda$ sufficiently close to the critical value $\lambda_c=0$, can be described by the stochastic differential equation (SDE):
\begin{equation}
    \label{eq:X}
dX_t = -(A(X_t-m)^2+\lambda)dt + \sigma dB_t,
\end{equation}
where $\mu=m+\sqrt{|\lambda |/A}$  is the stable fix point of the drift, $A$ is a time scale parameter, $B_t$ is a Brownian motion and $\sigma^2$ scales the variance. Disregarding the noise, this is the normal form of the co-dimension one saddle-node 
bifurcation~\cite{guckenheimer:1986} (supplementary text S5).
The square-root dependence of the stable state: $\mu-m\sim \sqrt{\lambda_c-\lambda}$ is the main signature of a saddle node bifurcation. It is observed for the AMOC shutdown in ocean only models as well as in coupled models, see Fig. \ref{fig:sn2}, in strong support of eq. (\ref{eq:X}) for the AMOC.

At time $t_0$, $\lambda(t)$ begins to change linearly towards $\lambda_c=\lambda(t_c)=0$:
\begin{equation}
\lambda(t)=\lambda_0(1-\Theta[t-t_0](t-t_0)/\tau_r),
\label{eq:lambda}
\end{equation}
where $\Theta[t]$ is the Heaviside function and $\tau_r=t_c-t_0 >0$ is the ramping time up to time $t_c$, where the transition eventually will occur. Time $t_c$ is denoted the tipping time, however, it can happen earlier due to a noise-induced tipping. As the transition is approached, the risk of a noise-induced tipping (n-tipping) prior to $t_c$ is increasing and at some point making the EWSs irrelevant for predicting the tipping. The probability for n-tipping can, in the small noise limit, be calculated in closed form, $P(t, \lambda)=1-\exp(-t/\tau_n(\lambda))$, with mean waiting time $\tau_n(\lambda) = (\pi/\sqrt{|\lambda|})\exp  (8|\lambda|^{\frac{3}{2}}/3\sigma^2)$ (supplementary text S4).

The mean and variance are calculated from the observations as the control parameter $\lambda(t)$ is possibly changing. These EWSs are inherently equilibrium concepts and statistical, thus a time-window, $T_{w}$, of a certain size is required for a reliable estimate. 
As the transition is approached, the differences between the EWSs and the pre-ramping values of the variance and autocorrelation (baseline) increase, thus, the shorter is the window $T_w$ required for detecting a difference.
Conversely, close to the transition critical slow down decreases the number of independent points within a window, thus calling for a larger window for a reliable detection.        
Within a short enough window, $[t-T_{w}/2, t+T_{w}/2]$, we may assume $\lambda(t)$ to be constant and the noise small enough so that the process \eqref{eq:X} for given $\lambda$ is well approximated by a linear SDE, the Ornstein-Uhlenbeck process~\cite{hasselmann:1976}. A Taylor expansion around the mean $\mu(\lambda)$ yields the approximation
\begin{equation}
    \label{eq:Xapprox}
    dX_t \approx -\alpha (\lambda) (X_t - \mu(\lambda)) dt + \sigma dB_t
\end{equation}
where $\mu (\lambda) = m + \sqrt{|\lambda|/A}$ and $\alpha (\lambda) = 2 \sqrt{|\lambda|/A}$ is the inverse correlation time. For fixed $\lambda$ the process is stationary, with mean $\mu$, variance $\gamma^2=\sigma^2/2\alpha$ and one-lag autocorrelation $\rho=\exp(-\alpha \Delta t)$ with step size $\Delta t = 1$ month. As $\lambda(t)$ increases, $\alpha$ decreases, and thus variance and autocorrelation increase. From $\mu$, $\gamma^2$ and $\rho$ the parameters of eq. \eqref{eq:X} are determined: $\alpha=-\log \rho/\Delta t$, $\sigma^2=2\alpha \gamma^2$, $A=\alpha/2(\mu-m)$ and $\lambda=(\sigma^2/4\gamma^2)^2/A$.  Closed form estimators for $\mu, \gamma^2$ and $\rho$ are obtained from the observed time series within a running window by maximum likelihood estimation (MLE) (supplementary text S1, see also \cite{DitlevsenLansky2020}).

The uncertainty is expressed through the variances of the estimators $\hat\gamma^2$ and $\hat\rho$ obtained from the observations within a time window $T_{w}$.   
Before the ramping where the process is stationary, the uncertainties can be made arbitrarily small by observing over a long time window. We may therefore assume that $\rho_0=\exp(-\alpha_0 \Delta t)$ and $\gamma^2_0 =\sigma^2/2\alpha_0$ are known, where $\alpha_0 = 2\sqrt{|\lambda_0|/A}$ and $\lambda_0$ is the baseline value before $t_0$. Detection of an EWS at some chosen confidence level $q$ (such as 95\% or 99\%) requires one of the estimates $\hat\gamma^2$ or $\hat\rho$ for a given window to be statistically different from the baseline values, which depends on the window size as well as how different the EWSs are from their baseline values.

\subsection*{Time scales in Early Warning Signals}

\begin{center}
\begin{figure}[htbp]
\begin{center}
\includegraphics[width=\textwidth]{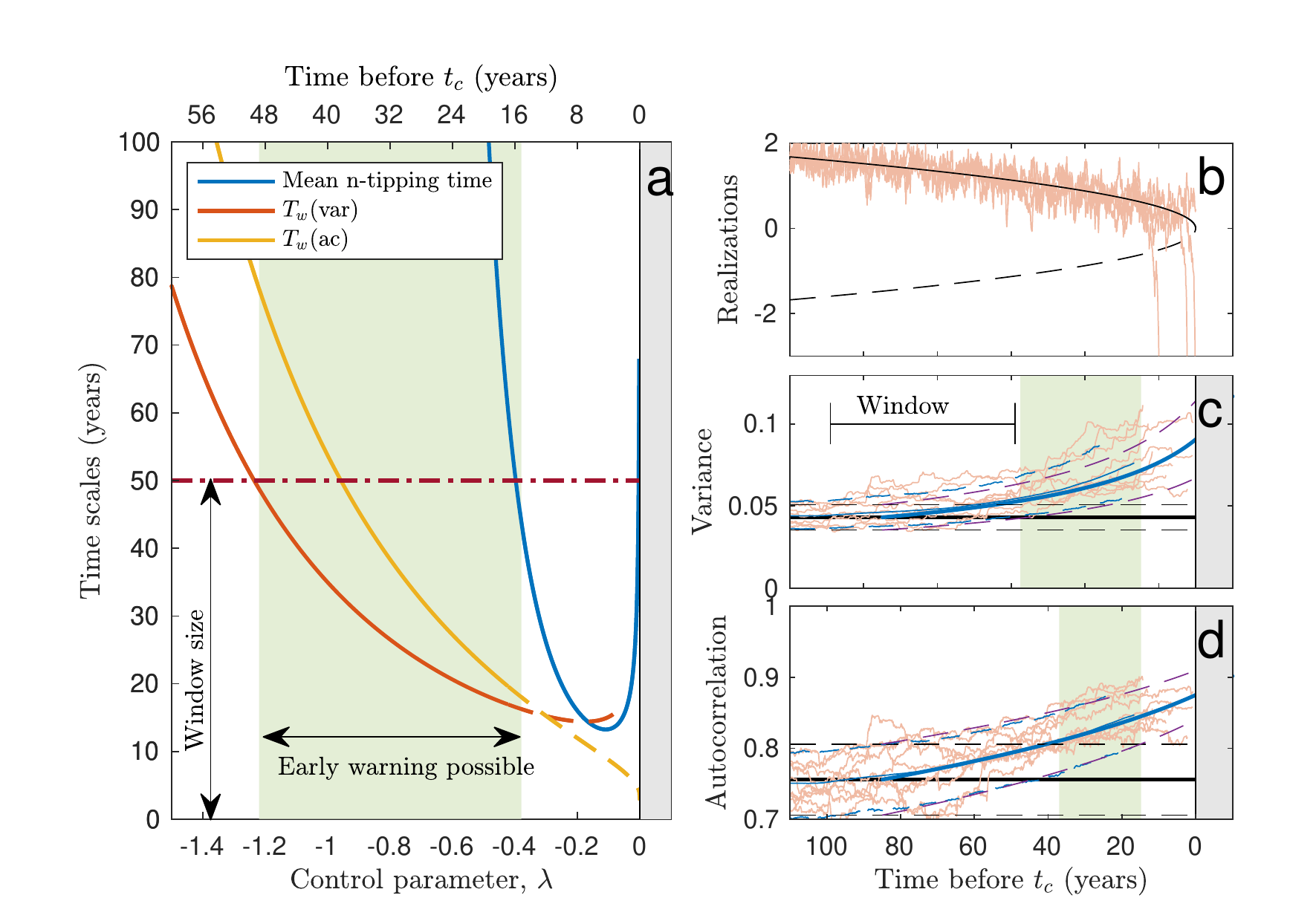}
\end{center}
\caption{\label{fig:simulations} Panel {\bf a} shows time scales involved in the critical transition ramping the control parameter $\lambda$ from $\lambda_0=-2.82$ to $\lambda_c=0$, with a ramping time $\tau_r=110$yrs and  $\sigma^2 = 0.29$. These parameters are obtained as best estimates from the HasISST data.  The time remaining before $t_c$ is shown on top of the plot. The red and orange curves shows the time window, $T_{w}$, needed in order to detect increase in variance (red) and autocorrelation (orange) above the pre-ramping values at the 95\% confidence level. Close to the bifurcation point, the (quasi-)stationarity approximation becomes less valid, which is indicated by the dashed part of the two curves. It is seen that detecting significant increase in autocorrelation requires a longer data window than detecting a significant increase in variance. With $T_{w}=50$yrs (red dot-dashed line) an increase in variance can only be detected at the 95\% confidence level after the red curve is below the 50yrs level.  
The blue curve shows the mean waiting time for a noise-induced transition, when this becomes shorter than the 50yrs level the EWS is no longer relevant, due to n-tipping occurring before $t_c$, thus the range of time, where an EWS can be applied is indicated by the green band (limited by the crossings of the red and blue curves with the size of the window). 
Panel {\bf b} shows ten model realizations of the ramped approach to $t_c$, notice a few n-tippings prior to $t_c$. The black (black dashed) curve is the stable (unstable) fixed point of the model. Panel {\bf c} shows the increased variance as EWS: Black line is the pre-ramping steady state value, while dashed lines are the two-sigma uncertainty range for calculating variance within the 50yr data window. The blue and dashed blue curves are the same, but for the model approaching the transition. The brown curves correspond to the ten realizations in Panel {\bf b}, while the green band corresponds to the green band in Panel {\bf a}. The thin blue lines are the same obtained from simulating 1000 realizations. Panel {\bf d} is the same as Panel {\bf c} but for the autocorrelation, where now the green band is narrower, corresponding to $T_{win}(ac)$ being smaller than the window size. 
}
\end{figure}
\end{center}

The detection of a forthcoming transition using statistical measures involves several time scales. The primary internal time scale is the autocorrelation time in the steady state. The period $\tau_r$ over which the control parameter changes from the steady state value to the critical value sets an external time scale. For given $\alpha (\lambda)$ and $q$-percentile the required time window $T_{w}(q, \alpha)$ to detect a change from baseline in EWSs at the given confidence level $q$ is given in closed form in the next section, (eq. \eqref{eq:Tvar} for variance and eq. \eqref{eq:Trho} for autocorrelation).
The involved time scales are summarized in Fig. \ref{fig:simulations}{\bf a}, where the required window size $T_{w}$ at the 95\% confidence level is plotted as a function of $\lambda$ for the variance (red curve) and autocorrelation (yellow curve). These are plotted together with the mean waiting time for n-tipping (blue curve). With $T_w= 50$yrs, increased variance can only be detected after the time when $\lambda(t)\approx -1.2$ (crossing of red and red-dashed curves). 
At that time a window of approximately 75yrs is required to detect an increase in autocorrelation, making variance the better EWS of the two. When $\lambda \approx -0.4$ the mean waiting time for n-tipping is smaller than the data window size. Thus, the increased variance can be used as a reliable EWS in the range  $-1.2 < \lambda(t) < -0.4$ indicated by the green band. How timely an early warning this is depends on the speed at which $\lambda(t)$ is changing from $\lambda_0$ to $\lambda_c$, i.e., the ramping time $\tau_r$. 
A set of 1000 realizations has been simulated with $\lambda_0=-2.82$ and $\tau_r=110$yrs, indicated by the time labels on top of Fig. \ref{fig:simulations}{\bf a}. Ten of these realizations are shown in Fig. \ref{fig:simulations}{\bf b} on top of the stable and unstable branches of fixed points of model \eqref{eq:X} (the bifurcation diagram). Fig. \ref{fig:simulations}{\bf c} ({\bf d}) shows the variance (autocorrelation) calculated from the realizations within a running 50yrs window (shown in Fig. \ref{fig:simulations}{\bf c}).   
The solid black line is the baseline value for $\lambda=\lambda_0$, while the solid blue line is the increasing value for $\lambda=\lambda(t)$. The calculated 95\% confidence level for the measurement of the EWS within the running window is shown by the dashed black and blue lines, respectively. The corresponding light blue curves are obtained numerically from the 1000 realizations. The green band in Fig. \ref{fig:simulations}{\bf c} corresponds to the green band in Fig. \ref{fig:simulations}{\bf a} and shows where early warning is possible in this case.

\subsection*{Statistics of Early Warning Signals}

The asymptotic variances of the estimators are Var$(\hat {\mu})
\approx
\gamma^2(1+\rho)/(1-\rho)n$, Var$(\hat \rho) \approx
(1-\rho^2)/n$ and Var$(\hat {\gamma}^2) \approx 2 (\gamma^2)^2(1+\rho^4)/(1-\rho^2)n$ (supplementary text S1). For $\alpha \Delta t\ll 1$ we approximate $(1+\rho^4)/(1-\rho^2) \approx 1/(\alpha \Delta t)$ and $1-\rho^2 \approx 2\alpha \Delta t$ and obtain 
\begin{eqnarray}
\label{eq:varvar}
\mbox{Var}(\hat {\gamma}^2) \approx \frac{2(\gamma^2)^2}{\alpha T_{w}} =\frac{\sigma^4}{2\alpha^3 T_{w}}; \quad \mbox{Var}(\hat \rho) \approx \frac{2\alpha \Delta t^2}{T_{w}},
\label{varestimates}
\end{eqnarray}
where $T_{w}= n\Delta t$ is the observation window.

The question is then how large $T_{w}$ needs to be to detect a statistically significant increase compared to the baseline values $\gamma^2_0$ and $\rho_0$. For a given estimate $\hat \gamma^2$, the estimated difference from the baseline variance is
\begin{eqnarray}
\label{eq:deltavar}
\Delta_{\gamma^2}=\hat \gamma^2-\gamma^2_0 &=& \gamma_0^2(\alpha_0/\hat\alpha-1),
\end{eqnarray}
and the estimated difference from the baseline autocorrelation is 
\begin{eqnarray}
\label{eq:deltarho}
\Delta_\rho=\hat \rho-\rho_0 &=& \rho_0(e^{(\alpha_0-\hat \alpha)\Delta t}-1) \approx \rho_0 (\alpha_0-\hat \alpha)\Delta t.
\end{eqnarray}
Since the two EWSs, $\gamma^2$ and $\rho$, are treated on an equal footing, in the following we let $\hat \psi$ denote either of the estimators \eqref{rhohat} or \eqref{gammahat}, the standard error is $s(\hat \psi)=\mbox{Var}(\hat \psi)^{1/2}$ (eq. \eqref{varestimates}) and $\hat \Delta$ denotes either of the two estimated differences \eqref{eq:deltavar} or \eqref{eq:deltarho}. The null hypothesis is that $\lambda = \lambda_0$, or equivalently $\alpha = \alpha_0$. The null distribution of $\hat \psi$ is assumed to be Gaussian (confirmed by simulations). A quantile $q$ from the standard Gaussian distribution expresses the acceptable uncertainty in measuring the statistical quantity $\psi$. We thus get that $\hat \Delta < q s(\hat \psi)$ at the $q$-confidence level (95\%, 99\% or similar) under the null hypothesis.  
To detect an EWS at the $q$-confidence level based on measuring $\psi$ at time $t$, we require that $\hat \Delta (t) > q (s(\hat \psi (t))+s(\psi_0))$, which, solved for $T_{w}$ gives for variance:
\begin{equation}
\label{eq:Tvar}
    T_{w}> 2q^2\left ( \frac{\alpha (t)/\sqrt{\alpha_0}+\alpha_0/\sqrt{\alpha(t)}}{\alpha_0 - \alpha(t) } \right )^2,
\end{equation}
and for autocorrelation,
\begin{equation}
\label{eq:Trho}
T_{w} > 2q^2\left ( \frac{\sqrt{\alpha_0}+\sqrt{\alpha(t)}}{\alpha_0 - \alpha(t) } \right )^2\rho_0^{-2}.
\end{equation} 
Substituting $\alpha (t) = 2\sqrt{A|\lambda(t)|}$, provides the time window $T_{w}$ needed to detect an EWS at time $t$ with large probability.

\subsection*{Predicting a forthcoming collapse of the AMOC}

\begin{center}
\begin{figure}[htbp]
\begin{center}
\includegraphics[width=12cm]{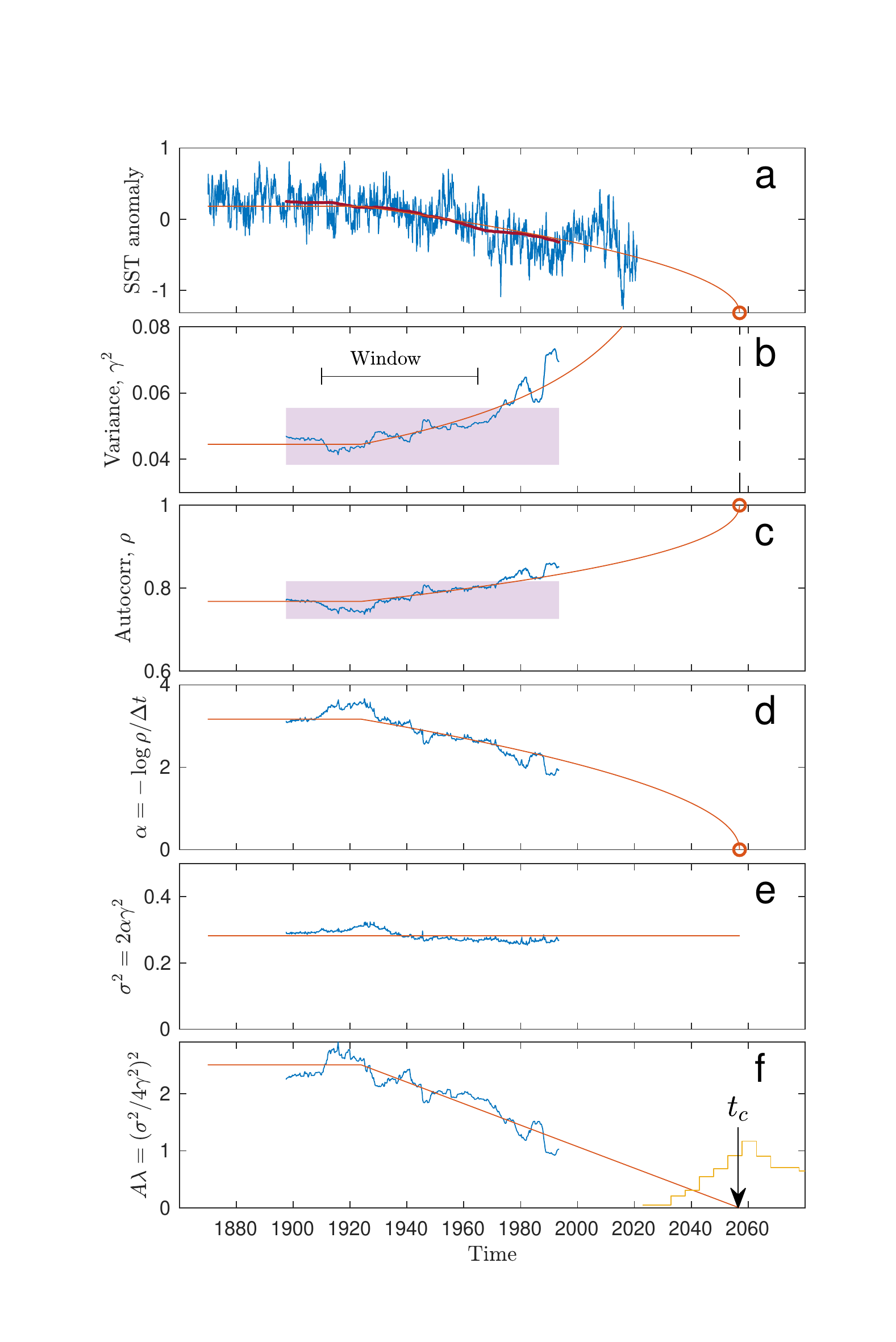}
\end{center}
\caption{\label{fig:dataanalysis} Panel {\bf a} shows the SST anomaly (identical to Figure 1{\bf e}) together with best estimate model of the steady state approaching a critical transition. Panels {\bf b} and {\bf c} show variance and autocorrelation calculated within running 50yr windows, similar to Figure \ref{fig:simulations}{\bf c} and {\bf d}. The two-sigma level (indicated by the purple band) is obtained using the model to estimate the time varying $\alpha$ (Panel {\bf d}) and $\sigma^2$ (Panel {\bf e}) from the data. Panel {\bf f} shows the best estimate for $t_c$. The yellow histogram is the probability density for $t_c$ obtained by maximum likelihood estimates (see Methods).
}
\end{figure}
\end{center}

The AMOC fingerprint shown in Fig. \ref{data}{\bf e} (replotted in Fig. \ref{fig:dataanalysis}{\bf a}) shows an increased variance, $\gamma^2$, and autocorrelation, $\rho$, plotted in Fig. \ref{fig:dataanalysis}{\bf b} and {\bf c} as functions of the mid-point of a 50yrs running window, i.e., the EWS obtained in 2020 is assigned to year 1995. The estimates leave the confidence band of the baseline values (pink area) around year 1970. This is not the estimate of $t_0$, which happened earlier and is still to be estimated; it is the year where EWSs are statistically different from baseline values. The estimates after 1970 stay consistently above the upper limit of the confidence interval and show an increasing trend, and we thus conclude that the system is approaching the tipping point with high probability. 

To estimate the tipping time once it has been established that the variance and autocorrelation are increasing, we use two independent methods to check the robustness of our results: 1. The first method is moment-based and uses the variance and autocorrelation estimates within the running windows. 2. The second method uses approximate MLE directly on model \eqref{eq:X}-\eqref{eq:lambda} with no running window. The advantage of the first method is that it has less model assumptions, however, it is sensitive to the choice of window size. The advantage of the second method is that it uses the information in the data more efficiently given model \eqref{eq:X}-\eqref{eq:lambda} is approximately correct, it has no need for a running window and does not assume stationarity after time $t_0$. In general, MLE is statistically the preferred method of choice giving the most accurate results with the lowest estimation variance. 

\begin{center}
\begin{figure}[htbp]
\begin{center}
\includegraphics[width=\textwidth]{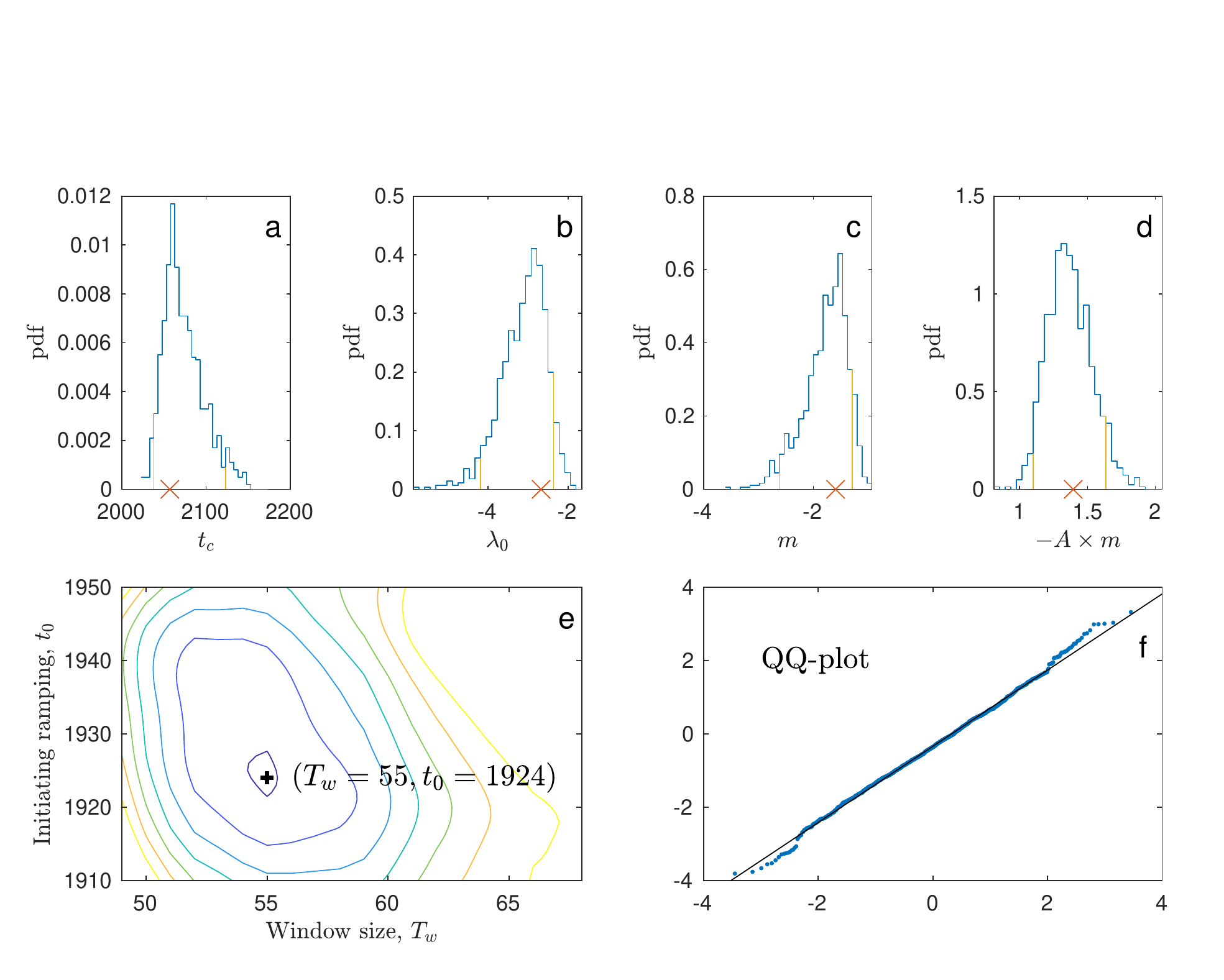}
\end{center}
\caption{\label{fig:pdf}With parameters obtained from the data, a set of 1000 realizations of the model are used in a bootstrap study to assess the uncertainty on parameters. Panels {\bf a}-{\bf d} are probability densities for $t_c$, $\lambda_0$, $m$ and $-A \times m$. Red crosses are the values obtained from the AMOC fingerprint data. The 95\% confidence intervals 
are indicated by orange lines. The critical time $t_c$ is 2057, and the 95 \% confidence interval is 2034-2128.
Panel {\bf e} shows the mean square error in fitting the ramping as a function of window size $T_w$ and time of initiating ramping, $t_0$. A unique minimum is found for $T_w=55$ yrs and $t_0=1924$. Panel {\bf d} shows the QQ-plot of residuals from the model, if points fall close to a straight line (black line) the model fits the data well. 
}
\end{figure}
\end{center}

\paragraph{1. Moment estimator of the tipping time} Within the running window, we obtain the parameters $\alpha(t)$ (Fig. \ref{fig:dataanalysis}{\bf d}) and $\sigma^2$ (Fig. \ref{fig:dataanalysis}{\bf e}) of the linearized dynamics, eq. \eqref{eq:Xapprox}. 
Then we obtain $A\lambda(t)$ from $\sigma^2$ and $\gamma^2(t)$ (Fig. \ref{fig:dataanalysis}{\bf f}), using that $A\lambda(t)=(\sigma^2/4\gamma^2(t))^2$. 
This is consistent with a linear ramping of $\lambda(t)$ beginning from a constant level $\lambda_0$ at a time $t_0$. By sweeping $t_0$ from 1910 to 1950 and $T_w$ from 45 to 65 yrs, we obtain $A\lambda_0$ and $\tau_r$ from least square error fit to the data. This shows a single minimum at $t_0=1924$ and $T_w = 55$yrs (Fig. \ref{fig:pdf}{\bf b}). Setting $t_0=1924$, we obtain $t_c$ from a linear fit (regressing $\lambda$ on $t$) from the crossing of the x-axis ($\lambda_c=0$). This is shown in Fig. \ref{fig:dataanalysis}{\bf f} (red line). 
This yields $-A\lambda_0=2.34$ year$^{-2}$ and $\tau_r=133$ years. Thus, the tipping time is estimated to be in year 2057, shown in Fig. \ref{fig:dataanalysis}{\bf f}.   
Since we have only obtained the combined quantity $A\lambda=(\sigma^2/4\gamma^2)^2$, we still need to determine $A$ and $m$ in Eq. \ref{eq:X}. We do that from the best linear fit to the mean level $\mu=m+\sqrt{|\lambda|/A}$ observing that 
$\mu=m+\sqrt{A|\lambda|}(1/A)=m+(\sigma^2/4\gamma^2)(1/A)$. The estimates are shown by the red curves in Fig. \ref{fig:dataanalysis}{\bf a}--{\bf f}. The red dot in Fig. \ref{fig:dataanalysis}{\bf a} is the tipping point and the dashed line in Fig. \ref{fig:dataanalysis}{\bf b} is the asymptote for the variance. 

With the parameter values completely determined, the confidence levels are calculated: The two-sigma levels around the baseline 
values of the EWS are shown by purple bands in Fig. \ref{fig:dataanalysis}{\bf b} and {\bf c}. Thus, both EWSs show increases beyond the two-sigma level from 1970 and onwards. 

\paragraph{2. Maximum likelihood estimator of the tipping time} We use approximate MLE on model \eqref{eq:X}--\eqref{eq:lambda}. The likelihood function is the product of transition densities between consecutive observations. However, the likelihood is not explicitly known for this model, and we therefore approximate the transition densities. From the data before time $t_0$ the approximation \eqref{eq:Xapprox} is used, where exact MLEs are available (supplementary text S1). This provides estimates of the parameters $\lambda_0, m$ as a function of parameter $A$, as well as the variance parameter $\sigma^2$.

To estimate $A$ and $\tau_r$, the observations after time $t_0$ are used. After time $t_0$, the linear approximation \eqref{eq:Xapprox} is no longer valid, because the dynamics are approaching the bifurcation point and the non-linear dynamics will be increasingly dominating. The likelihood function is the product of transition densities, which we approximate with a numerical scheme, the Strang splitting, which has shown to have desirable statistical properties for highly non-linear models, where other schemes, such as the Euler-Maruyama approximation is too inaccurate \cite{pilipovic2022} (supplementary text S2). 

The optimal fit is $t_0 = 1927$ and $t_c = 2069$ with a 95\% confidence interval 2034-2128 obtained by bootstrap (see below). These estimates are close to the estimates obtained by the moment method.

\paragraph{Uncertainty in the estimate of the tipping time}
The likelihood approach provides asymptotic confidence intervals, however, these assume that the likelihood is the true likelihood. To incorporate also the uncertainty due to the data generating mechanism \eqref{eq:X} not being equal to the Ornstein-Uhlenbeck process \eqref{eq:Xapprox} used in the likelihood, we chose to construct parametric bootstrap confidence intervals. This was obtained by simulating 1000 trajectories from the original model with the estimated parameters, and repeat the estimation procedure on each data set. Empirical confidence intervals were then extracted from the 1000 parameter estimates. These were indeed larger than the asymptotic confidence intervals provided by the likelihood approach, however, not by much.

From the tipping times estimated on each simulated data set, the probability density function (PDF) (Fig. \ref{fig:dataanalysis}{\bf f}, yellow histogram) is obtained. The median is $\langle t_c\rangle=2066$ and the 95\% confidence interval is $2034-2128$. The small discrepancy in median is probably due to the approximate model used for estimation being different from the data generating model \eqref{eq:X}, confirming that the linear model still provides valid estimates even if the true dynamics are unknown. To test the goodness-of-fit, normal residuals (supplementary text S3) were calculated for the data. These are plotted in Fig. \ref{fig:pdf}{\bf f} as a quantile-quantile plot. If the model is correct, the points fall close to a straight line. The model is seen to fit the data well, further supporting the obtained estimates.

\subsection*{Summary}
We have provided a novel robust statistical analysis to quantify the uncertainty in observed EWSs for a forthcoming critical transition. The confidence depends on how rapid the system is approaching the tipping point. With this the significance of the observed EWSs for the AMOC has been established. This is a stronger result than just observing a significant trend in the EWS, by, say, a Kendall's $\tau$ test. Here we calculate when the EWS are significantly above the natural variations. Furthermore, we have provided a method to not only determine whether a critical transition will happen, but also an estimate of when it will happen. We predict with high confidence the tipping to happen as soon as 2057. This is indeed a worrisome result, which should call for fast and effective measures to reduce global greenhouse gas emissions in order to avoid the steadily change of the control parameter towards the collapse of the AMOC (i.e. reduce temperature increase and fresh water input through ice melting into the North Atlantic region). As a collapse of the AMOC has strong societal implications~\cite{kemp:2022}, it is important to monitor the flow and EWS from direct measurements~\cite{baehr:2007,sevellec:2018,alexander-turner:2018}.

\subsection*{Acknowledgements}
This work has received funding under the project Tipping Points in the Earth System (TiPES) from the European Union’s Horizon 2020 research and innovation programme under grant agreement no. 820970. This is TiPES contribution \#214. SD received funding from Novo Nordisk Foundation NNF20OC0062958.

\bibliographystyle{nature}
\bibliography{climate}

\newpage

\appendix

\section*{Supplementary text}

{
\subsection*{S1 Maximum likelihood estimators of the Ornstein-Uhlenbeck process} \label{Appendix:OUMLE}
}

{
To obtain eq. \eqref{eq:varvar} we need the maximum likelihood estimator (MLE) of the approximate model.
The approximate model is an Ornstein-Uhlenbeck (OU) process, defined as the solution to the equation
\begin{equation}
    \label{eq:OU}
    dX_t = -\alpha (X_t - \mu) dt + \sigma dB_t.
\end{equation}
This is a Gaussian process with well-known properties \cite{DitlevsenLansky2020,FormanSorensen2008}.
The variance is $\gamma^2 = \sigma^2/2\alpha$ and the $\Delta t$-lag autocorrelation is $\rho = e^{-\alpha \Delta t}$. The likelihood function of the parameters given observations $(x_0, x_1, \ldots , x_n)$ is the product of the transition densities
\begin{equation} 
\label{eq:Ln}
L_{n} ( \theta ) = \prod_{i=1}^{n} p(\triangle,x_{i-1},x_{i};
\theta)
\end{equation}
where $\theta = (\mu, \rho, \gamma^2)$. Here, $x_i = x(t_i)$ and $\Delta t = t_i - t_{i-1}$. 
The transition density is normal with conditional mean $E(X_i | X_{i-1} = x_{i-1}) = x_{i-1} \rho + \mu (1-\rho)$ and conditional 
variance $\gamma^2 (1-\rho^2)$, 
\begin{equation}
\label{eq:transitiondensity}
p(\triangle,x_{i-1},x_{i}; \theta) = \dfrac
{1} {\sqrt{2\pi\gamma^{2} (1 - \rho^2)} } \exp \left ( - \dfrac {
  (x_{i}- x_{i-1} \rho - \mu (1- \rho))^{2} } {2\gamma^{2}
(1-\rho^2)} \right ),
\end{equation}
see \cite{DitlevsenLansky2020,FormanSorensen2008} for details. The likelihood function is the joint probability of the observed data viewed as a function of the parameters of the statistical model, in this case discrete observations from the Ornstein-Uhlenbeck process. Considering the observed sample as fixed, the likelihood is a function of the parameters. The likelihood principle states that all the information about the parameter $\theta$ is given in the likelihood function. The maximum likelihood estimator is the value of $\theta$ which maximizes the probability of observing the given sample. In practice, the maximum of the likelihood function is found by taking the derivative with respect to the parameters (the score) and equate it to zero (the likelihood equation). For further details about likelihood theory, see any textbook in mathematical statistics. 

The maximum likelihood estimators (MLEs) derived from eqs. \eqref{eq:Ln} and \eqref{eq:transitiondensity} are
\begin{eqnarray}
\label{muhat} \hat {\mu} &=& \frac{1}{n} 
\sum_{i=1}^{n} x_{i} + \dfrac{\hat \rho}
{n(1-\hat \rho)}(x_{n} - x_{0})  \approx \frac{1}{n+1} 
\sum_{i=0}^{n} x_{i} \equiv \bar x, \\
\label{rhohat} \hat \rho  &=&
\dfrac{\sum_{i=1}^{n}
(x_{i}-\hat {\mu}) (x_{i-1}-\hat {\mu})} {\sum_{i=1}^{n}
(x_{i-1}-\hat {\mu})^{2} } ,\\
\label{gammahat} \hat {\gamma}^2 &=& \dfrac{{\sum_{i=1}^n 
  \left ( x_i -x_{i-1}\hat \rho - \hat \mu
    (1-\hat \rho )\right )^2 }}{{n \left ( 1- \hat \rho^2\right )
}},
\end{eqnarray}
the symbol \,
$\hat{}$ \, indicates an estimator. These are obtained as follows. The score function is the vector of derivatives of the log-likelihood
function with
respect to the parameters. The MLE is given as solution to the
likelihood equations $\partial_{\theta_k} \log L_{n} ( \theta ) =0$,
where $\theta_k$ is either $\mu, \rho$ or $\gamma^2$. The
score function is
\begin{eqnarray*}
\frac{\partial}{\partial \mu} \log L_{n}(\theta) &=&
\dfrac{(1-\rho)}{\gamma^{2}  (1-\rho^2)} \sum_{i=1}^{n} (x_{i}- x_{i-1} \rho -\mu (1-\rho)), \\
\frac{\partial}{\partial \rho} \log L_{n}(\theta) &=&
\dfrac{n\rho}{1-\rho^2} +  \dfrac {\sum_{i=1}^{n}
(x_{i} - x_{i-1} \rho -\mu (1-\rho))  (x_{i-1} - \mu ) } {\gamma^{2} (1-\rho^2)} \\
&&- \dfrac {\rho\sum_{i=1}^{n} (x_{i}
- x_{i-1} \rho -\mu (1-\rho))^{2} }
{\gamma^{2} (1-\rho^2)^{2}},\\
\frac{\partial}{\partial \gamma^2} \log L_{n}(\theta) &=&
-\dfrac{n}{2\gamma^2} + \dfrac {\sum_{i=1}^{n} (x_{i}- x_{i-1} \rho
  -\mu (1-\rho))^{2} } {2\gamma^{4}
(1-\rho^2)},
\end{eqnarray*}
whose zeros provide the MLEs in equations
\eqref{muhat}--\eqref{gammahat}. 
It requires that $\sum_{i=1}^{n} (x_{i}-\hat{\mu})
(x_{i-1}-\hat{\mu}) > 0$, otherwise the MLE does not exist.
}

{
 The Fisher Information $\mathcal{I}$ of the MLEs equals minus the expectation of the
Hessian $\mathcal{H}$ of the log-likelihood function. For the OU log-likelihood, the
elements of $\mathcal{H}$ are given by
\begin{eqnarray*}
\frac{\partial^2}{\partial \mu^2} \log L_{n}(\theta) &=&
-\frac{n(1-\rho)}{\gamma^2 (1+\rho)},\\
\frac{\partial^2}{\partial \mu \rho} \log L_{n}(\theta) &=&
\sum_{i=1}^n  \left
( C_1 (x_{i-1}-\mu ) +C_2 (x_i - x_{i-1}\rho -\mu (1-\rho) ) \right ) ,\\
\frac{\partial^2}{\partial \mu \gamma^2} \log L_{n}(\theta) &=& C_3 \sum_{i=1}^{n} (x_{i}- x_{i-1} \rho   -\mu (1-\rho)),\\
\frac{\partial^2}{\partial \rho^2} \log L_{n}(\theta) &=& \frac{n(1+\rho^2)}{(1-\rho^2)^2}
  + C_4 \sum_{i=1}^{n}
(x_{i} - x_{i-1} \rho -\mu (1-\rho))  (x_{i-1} - \mu ) -
            \frac{1}{\gamma^2 (1-\rho^2)}
    \sum_{i=1}^{n}(x_{i-1}-\mu)^2
  \\
&& -\frac{1+3 \rho^2}{\gamma^2 (1-\rho^2)^3} \sum_{i=1}^{n} (x_{i} - x_{i-1} \rho -\mu (1-\rho))^{2},\\
\frac{\partial^2}{\partial \rho \gamma^2} \log L_{n}(\theta) &=&  C_5\sum_{i=1}^{n}
(x_{i} - x_{i-1} \rho -\mu (1-\rho))  (x_{i-1} - \mu ) 
+  \frac{\rho}{\gamma^4(1-\rho^2)^2} \sum_{i=1}^{n} (x_{i} - x_{i-1} \rho -\mu (1-\rho))^{2},\\
\frac{\partial^2}{\partial (\gamma^2)^2 } \log L_{n}(\theta) &=& \dfrac{n}{2\gamma^4} - \dfrac {\sum_{i=1}^{n} (x_{i}- x_{i-1} \rho
  -\mu (1-\rho))^{2} } {\gamma^{6}
(1-\rho^2)},
\end{eqnarray*}
where $C_i, i=1, \ldots, 5,$ are deterministic constants that will
disappear when taking expectations. Using that $E(X_{i}-\mu)^2=\gamma^2$, $E(X_{i} - X_{i-1} \rho -\mu (1-\rho))^{2}=\gamma^2 (1-\rho^2)$ and $E(X_{i} - X_{i-1} \rho -\mu (1-\rho))  (Y_{i-1} - \mu )=0$,  we obtain the Fisher Information 
$$\mathcal{I} = -E \mathcal{H} =n\begin{bmatrix} 
\frac{(1-\rho)}{\gamma^2 (1+\rho)} & 0&0 \\
0 & \frac{1+\rho^4}{(1-\rho^2)^2} & \frac{\rho}{\gamma^2(1-\rho^2)} 
\\
0 &\frac{\rho}{\gamma^2(1-\rho^2)} & \frac{1}{2 \gamma^4 }
\end{bmatrix}.
$$
The inverse of the Fisher Information provides the asymptotic
covariance matrix,
$$\frac1n \begin{bmatrix} 
\frac{\gamma^2 (1+\rho)}{(1-\rho)} & 0&0 \\
0 & 1-\rho^2 & 2\rho \gamma^2 
\\
0 &2\rho \gamma^2 &  \frac{2\gamma^4(1+\rho^4)}{1-\rho^2} 
\end{bmatrix}.
$$
The diagonal elements provide the asymptotic variances of $\mu,
\rho$ and $\gamma^2$, respectively. 
}

{
\subsection*{S2 Estimator of the tipping time}
\label{Appendix:Strang}

The process is given as solution to 
\begin{eqnarray}
\label{X2A}
dX_t &=& -(A(X_t-m)^2 +\lambda_t)dt + \sigma dB_t,\\
    \label{eq:lambdaA}
    \lambda_t &=& \lambda_0 (1- \Theta [t - t_0] (t-t_0)/\tau_r).
\end{eqnarray}
and we wish to estimate the parameters $\theta = (A, m , \lambda_0, \tau_r, \sigma)$ from observations $(x_0, x_1, \ldots , x_n)$ before time $t_0$ and observations $(y_0, y_1, \ldots , y_n)$ after time $t_0$, of process $X_t$ defined by \eqref{X2A}. 
This equation
cannot be explicitly solved, and the exact distribution 
is not explicitly known. A standard way to solve this is approximating the transition density by a Gaussian distribution obtained by the Euler-Maruyama scheme. However, the estimators obtained from the Euler-Maruyama pseudo-likelihood are known to be biased, especially in non-linear models \cite{pilipovic2022}. Instead we use a two-step procedure: First we estimate $\alpha_0 = 2\sqrt{A |\lambda_0|}$, $\mu_0 = m+\sqrt{|\lambda_0|/A}$ and $\sigma^2$ from the stationary part before time $t_0$, using estimators \eqref{muhat} -- \eqref{gammahat}, where $\alpha_0 = -\log (\rho)/\Delta t$ and $\sigma^2 = 2 \alpha_0  \gamma^2$. This yields estimates $\lambda_0 (A) = -\alpha_0^2/4A$ and $m(A)= \mu_0-\alpha_0/2A$ as a function of parameter A and the estimated parameters. The two remaining parameters $A$ and $\tau_r$ are then estimated from the data after time $t_0$, where we no longer can use the OU process, since the linear approximation breaks down when the tipping point is approached. Simplifying by assuming that $\lambda$ is constant between observations, i.e., piecewise constant and jumping every month where new AMOC observations are available, we obtain transition densities that are non-linear transformations of Gaussian densities, making the inference problem tractable as follows. We use a pseudo-likelihood induced by the Strang splitting scheme, shown to be robust for highly non-linear models \cite{pilipovic2022}. Consider the two subsystems
\begin{eqnarray}
dX_t^{(1)} &=& -\alpha (\lambda)(X_t^{(1)} - \mu (\lambda)) dt + \sigma dB_t,\label{eq:LT1} \\
dX_t^{(2)} &=& -A(X_t^{(2)}-\mu (\lambda))^2dt, \label{eq:LT2}
\end{eqnarray}
where $\alpha (\lambda) = 2\sqrt{A |\lambda|}$ and $\mu (\lambda) = m+\sqrt{|\lambda|/A}$. The drift of subsystem \eqref{eq:LT1} is the Taylor expansion of the drift in eq. \eqref{X2A} to first order around the fixed point $\mu (\lambda)$ and is an OU process, of which we know
the distribution and the likelihood (see S1). Eq. \eqref{eq:LT2} is a deterministic
equation with the non-linear part, which solution is also known. We obtain the following two flows:
\begin{eqnarray*}
\phi^{(1)}_{\Delta}(x) \, := \, (X_{t+\Delta}^{(1)} \mid X_{t}^{(1)} = x)&=& x e^{-\alpha (\lambda ) \Delta } + \mu(\lambda)(1-e^{ -\alpha (\lambda )} )+\xi_t\\
\phi^{(2)}_{\Delta}(x) \, := \,(X_{t+\Delta}^{(2)} \mid X_{t}^{(2)} = x)&=& 
\frac{\mu (\lambda ) A \Delta (x-\mu (\lambda )) + x}{A \Delta (x-\mu (\lambda )) + 1} 
\end{eqnarray*}
where $\xi_t \sim N(0,\Omega_{\Delta})$,  $\Omega_{\Delta} = \frac{\sigma^2}{2\alpha (\lambda )}(1-e^{-2\alpha (\lambda ) \Delta })$.

The Strang splitting \cite{pilipovic2022} then approximates by
\begin{equation}
\label{eq:strang}
(X_{t+\Delta} \mid X_{t} = x)= \left ( \phi^{(2)}_{\Delta/2} \circ \phi^{(1)}_{\Delta} \circ  \phi^{(2)}_{\Delta/2}\right ) (x) = \phi^{(2)}_{\Delta/2} \left ( e^{-\alpha (\lambda_t) \Delta} \phi^{(2)}_{\Delta/2}(x) + \mu (\lambda_t) (1-e^{-\alpha (\lambda_t) \Delta})+ \xi_t\right ),
\end{equation}
which is defined for all $x> \mu(\lambda_t) -2/A\Delta$. Since we are only interested in simulating the process up to the time where $X_t$ crosses the separatrix between the two attractors, which happens for $x < m$, we require that $m> m + \sqrt{|\lambda_t|/A}-2/A\Delta \geq m + \sqrt{|\lambda_0|/A}-2/A\Delta$, i.e.,  $\Delta < 2/  \sqrt{A|\lambda_0|}= 4/\alpha_0$. This is always fulfilled, since $\Delta = 1/12$ and $\alpha_0$ is estimated to be less than 4.

The transition density \eqref{eq:strang} is a nonlinear transformation of a Gaussian random variable, leading to the pseudo-loglikelihood function (up to a constant)
\begin{equation}
\label{eq:strangloglik}
    -\log L_n (A,\tau_r) = \frac{1}{2} \sum_{i=1}^n\log (\Omega_{\Delta})+\sum_{i=1}^n \frac{Z_i^2}{2\Omega_{\Delta}}  -\sum_{i=1}^n\log |\frac{d}{dx}(\phi^{(2)}_{\Delta/2})^{-1} (y_i)|
\end{equation}
where
$$Z_i = (\phi^{(2)}_{\Delta/2})^{-1} (y_i) - e^{-\alpha (\lambda_{t_{i-1}}) \Delta} \phi^{(2)}_{\Delta/2}(y_{i-1}) + \mu (\lambda_{t_{i-1}}) (1-e^{-\alpha (\lambda_{t_{i-1}}) \Delta}),$$
see \cite{pilipovic2022} for details. The first two terms in \eqref{eq:strangloglik} are the standard terms from a Gaussian likelihood, the last term originates from the non-linear transformation. Estimates of parameters $A, \tau_r$ are then obtained by minimizing \eqref{eq:strangloglik}. Since division by $A$ enters the calculations of $\lambda_0$ and $m$ and thus the pseudo-likelihood, estimates are sensitive to small values of $A$. We therefore regularize the optimization problem by adding a penalization term on small values of $A$. The term $-pn(1/A - 1)$ is added to \eqref{eq:strangloglik} for $A < 1$, where $p \geq 0$ is a penalization parameter determined by cross-validation on simulated data sets by minimizing the mean squared distance between the estimated ramping time on each data set to the value of the ramping time used in the simulation. The optimal value was $p = 0.004$.

The parameter estimates are found numerically by minimizing $-\log L_x(\theta)$. For this we apply the optimizer {\tt optim} in R, using the Nelder-Mead algorithm. 

Confidence intervals are obtained by parametric bootstrap: 1000 repetitions of the model are simulated with the estimated parameters, and on each synthetic data set, parameters are estimated. The empirical quantiles of the 1000 estimates thus obtained are used to construct confidence intervals.
}

\subsection*{S3 Model control}

To test the model fit, uniform residuals, $u_i, i = 1, \ldots , n$, were calculated for the AMOC data using the estimated parameters from the MLE method as follows. The model assumes that observation $x_i$ follows some distribution function $F_{i,\hat \theta}$ for the estimated parameter values $\hat \theta$. If this is true, then $u_i = F_{i,\hat \theta} (x_i)$ is uniformly distributed on $(0,1)$. Transforming these residuals back to a standard normal distribution provides standard normally distributed residuals if the model is true. Thus, a normal quantile-quantile plot reveals the model fit. The points should fall close to a straight line. The reason for making the detour around the uniform residuals is twofold. First, since the data is not stationary, each observation follows its own distribution, and residuals cannot be directly combined. Second, since the model is stochastic, standard residuals are not well-defined, and observations should be evaluated according to their entire distribution, not only the distance to the mean.  

\subsection*{S4 Noise induced tipping}

The drift term in eq. \eqref{eq:X} is the negative gradient of a potential, $f(x, \lambda) = -\partial_x V(x,\lambda) = -(A(x-m)^2 + \lambda)$ with $V(x,\lambda) = A(x-m)^3/3+(x-m)\lambda$. For $\lambda < 0$, the drift has two fixed points, $m\pm \sqrt{|\lambda|/A}$. The point $m+\sqrt{|\lambda|/A}$ is a local minimum of the potential $V(x,\lambda)$ and is stable, whereas $m-\sqrt{|\lambda|/A}$ is a local maximum and unstable. The system thus has two basins of attraction separated by $m-\sqrt{|\lambda|/A}$, with a drift towards either $m+\sqrt{|\lambda|/A}$ or $-\infty$ dependent on whether $X_t > m-\sqrt{|\lambda|/A}$ or $X_t < m-\sqrt{|\lambda|/A}$. We denote the two basins of attraction the normal and the tipped state, respectively. When $\lambda = 0$, the normal state disappears and the system undergoes a bifurcation and $X_t$ will be drawn towards $-\infty$.

Due to the noise, the process can escape into the tipped state by crossing over the potential barrier  $\Delta(\lambda)= V(-\sqrt{|\lambda|}, \lambda) - V(\sqrt{|\lambda|}, \lambda) = 4 |\lambda|^{\frac{3}{2}}/3A^{\frac12}$.
Assume $X_t$ to be close to $m+\sqrt{|\lambda|/A}$ at some time $t$, i.e., in the normal state. 
The escape time will asymptotically (for $\sigma \rightarrow 0$) follow an exponential distribution such that
\begin{equation}
    \label{eq:il}
P(t, \lambda)=1-\exp(-t/\tau_n(\lambda))
\end{equation}
where $P(t, \lambda)$ is the probability of observing an escape time shorter than $t$ for a given value of $\lambda$. The mean noise induced escape time $\tau_n(\lambda)$ is \cite{Berglund2013, WentzellFreidlin}:

\begin{equation}
\tau_n(\lambda) = \frac{2\pi \exp(2\Delta(\lambda)/\sigma^2)}{ \sqrt{V''(m+\sqrt{|\lambda|/A},\lambda)|V''(m-\sqrt{|\lambda|/A},\lambda)|} }=(\pi/\sqrt{A|\lambda|})\exp  (8|\lambda|^{\frac{3}{2}}/3A^{\frac12}\sigma^2).
\label{eq:tau}
\end{equation}

Assume that the rate of change of $\lambda (t)$ follows eq. \eqref{eq:lambda}, then for $\tau_r < \tau_n(\lambda)$, the waiting time for a random crossing is so long that a crossing will not happen before a bifurcation induced transition happens (b-tipping). If $\tau_r > \tau_n(\lambda)$, a noise-induced tipping is expected before the bifurcation point is reached. Since $\tau_n(\lambda)$ decreases with increasing $\lambda$, at some point, the two time scales will end up matching.

\subsection*{S5 Normal form of the saddle-node bifurcation}
\begin{center}
\begin{figure}[htbp]
\begin{center}
\includegraphics[width=\textwidth]{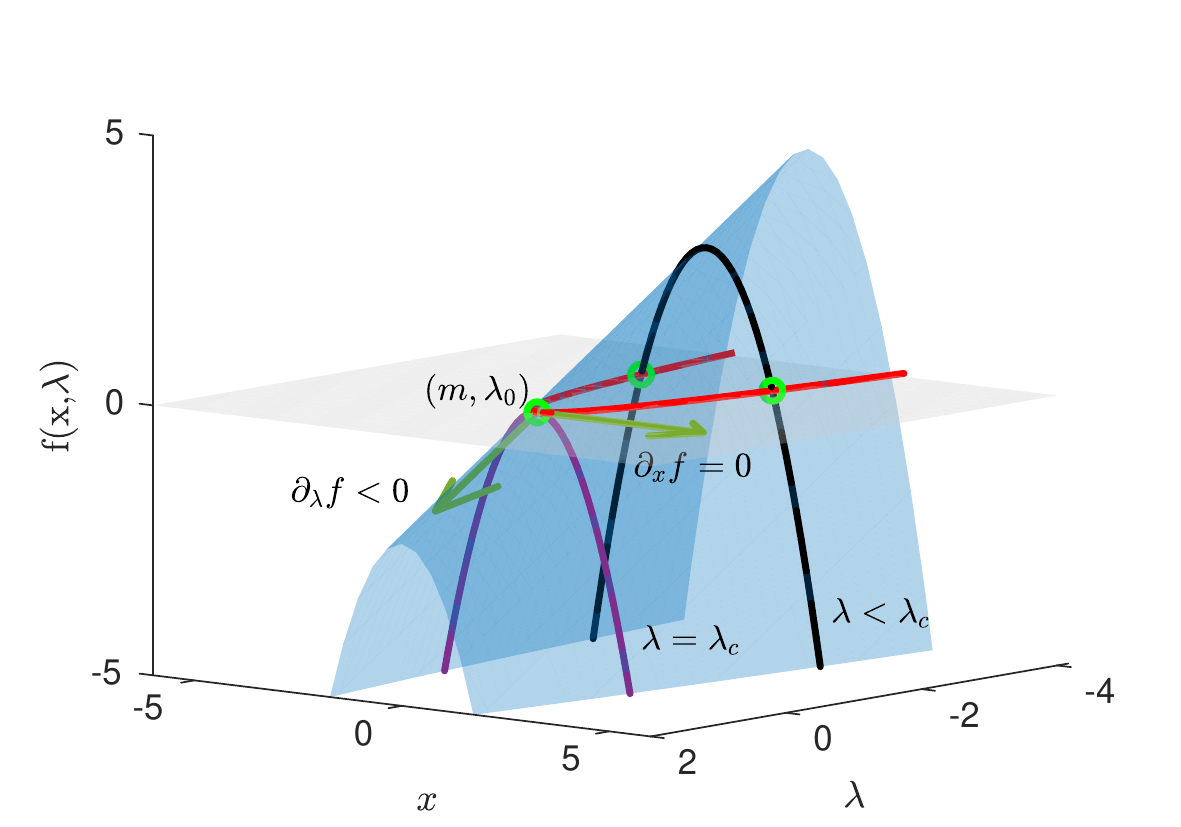}
\end{center}
\caption{\label{fig:sn} 
The function $f(x, \lambda)$ near a saddle point where a stable and an unstable fix point merge at a saddle-node bifurcation. For $\lambda<\lambda_c $ there are two fix points (green) where the black curve pass through the null-plane, $f=0$ (grey). The point in front is the stable fix point, while the point in the back is the unstable fix point. The red curve of fix points is the bifurcation curves, with the stable branch in front and unstable branch in the back.  The purple curve is $f(x,\lambda_c)$ which touch the null-plane in one point $(x_0,\lambda_c)$.
At this point it is seen that  $\partial_x f =0$ and $\partial_\lambda f < 0$, indicated by the dark green tangents to the surface. }
\end{figure}
\end{center}

Consider the general dynamical equation

\begin{equation}
    \frac{dx}{dt}=f(x,\lambda),
\end{equation}
where $x$ is a variable and $\lambda$ is a (fixed) parameter. A point $x_0$ with $f(x_0,\lambda)=0$ is a fix point or steady state. A fix point is stable/unstable if $\partial_xf(x,\lambda)_{x=x_0}$ is negative/positive, thus the fix point is attracting/repelling. If $f(x,\lambda)$ is not a linear function of $x$, multiple steady states may exist. A saddle-node bifurcation occur when changing the control parameter $\lambda$ through a critical value $\lambda_c$ a stable and an unstable fix point merge and disappears. The situation is shown in the figure, where the blue surface is $f(x,\lambda)$, while the grey (null-) plane is $f(x,\lambda)=0$. For a constant value of $\lambda$ the dynamics is determined by the black curve. The fix points are determined by the intersection with the null-plane (green), the point in the front is the stable fix point, while the further point is the unstable fix point. When changing $\lambda$ towards $\lambda_c=0$, the two fix points merge at the saddle-node bifurcation $(m,\lambda_c)$ (green).
The normal form of the saddle-node is obtained by expanding $f(x,\lambda)$ to lowest order around the point $(m,\lambda_c)$, noting that $f(m,\lambda_c)=0$, $\partial_x f(x,\lambda)_{(x,\lambda)=(m,\lambda_c)} =0$ and $\partial_\lambda f(x,\lambda)_{(x,\lambda)=(m,\lambda_c)} < 0$ (see Fig \ref{fig:sn}):

\begin{equation}
    f(x,\lambda)\approx \frac{1}{2} \partial^2_{x^2}f(x,\lambda)_{(x,\lambda)=(m,\lambda_c)}\times (x-m)^2+ \partial_\lambda f(x,\lambda)_{(x,\lambda)=(m,\lambda_c)} \times (\lambda-\lambda_c) =-A (x-m)^2- \tilde{\lambda},
\end{equation}
where $A=-\frac{1}{2} \partial^2_{x^2}f(x,\lambda)_{(x,\lambda)=(m,\lambda_c)}$ and $\tilde{\lambda}=-\partial_\lambda f(x,\lambda)_{(x,\lambda)=(m,\lambda_c)}\times (\lambda-\lambda_c)$.
This is the normal form for the saddle-node bifurcation.
Thus, close to the bifurcation point the stable steady state is 

\begin{equation}
 x_0=m+\sqrt{-\tilde{\lambda}/A}.
 \label{ss}
\end{equation} 

In order to see that this is indeed the case for the AMOC transition also in comprehensive climate models, Fig. \ref{fig:sn2} 
is adapted from the model intercomparison study\cite{rahmstorf:2005}. The steady state curves obtained are from simulations, with a very slowly changing control parameter (freshwater forcing). Top panel shows ocean only models, while bottom panel shows atmosphere-ocean models. The curves are, even away from the transition, surprisingly well fitted by eq. \eqref{ss}. Note that for some models the transition happens before the critical point, as should be expected from noise induced transitions. Note also that the data has been smoothed such that increasing variance close to the transition is not visible. This observation strongly supports the assumption of a saddle-node bifurcation, while it also shows that $(m, \lambda_c)$ (black dots) are quite different between models, thus calls for reliable determination from observations.

\subsection*{S6 The AMOC proxy}
\begin{center}
\begin{figure}[htbp]
\begin{center}
\includegraphics[width=\textwidth]{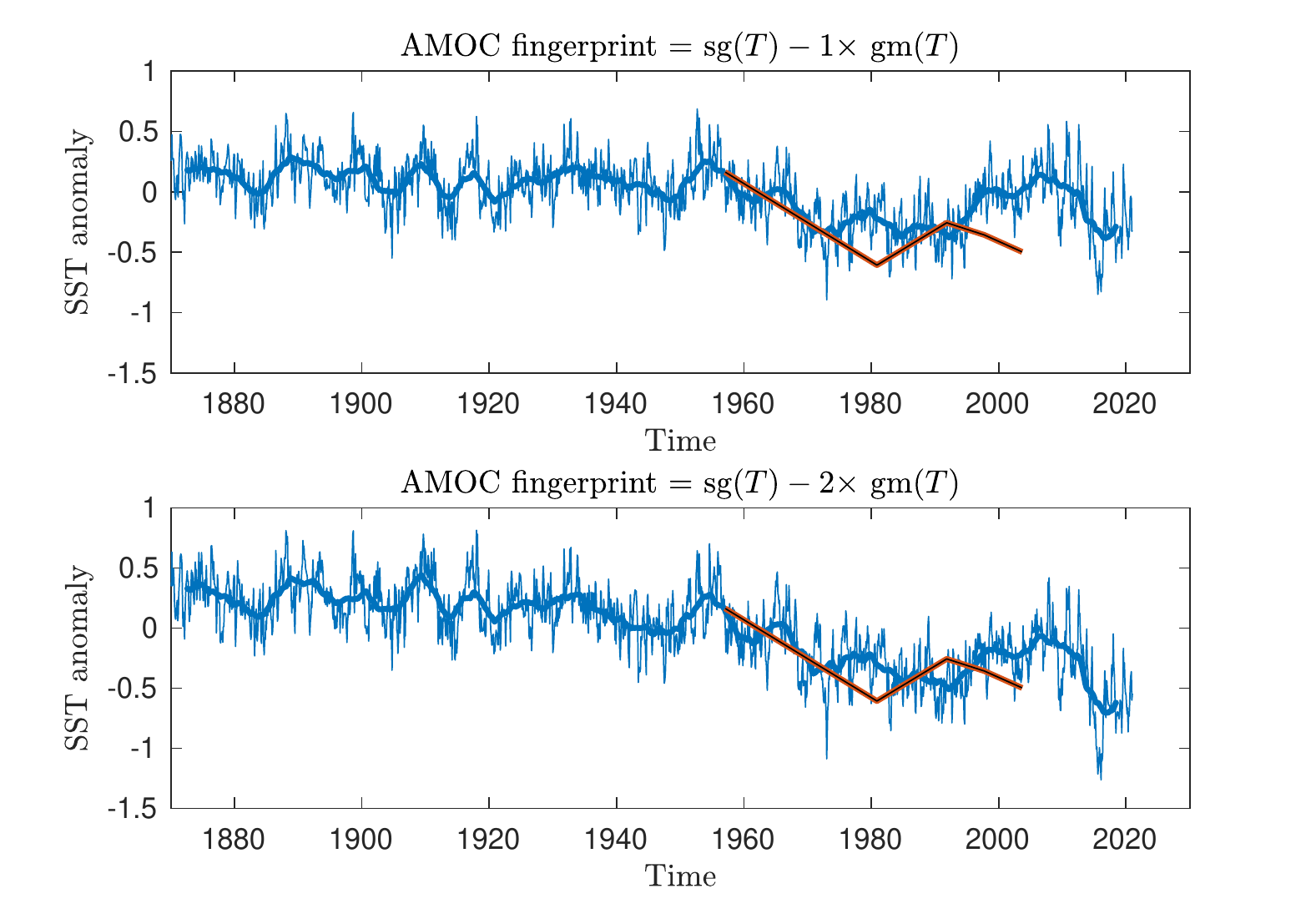}
\end{center}
\caption{\label{fig:calibrated} 
In the SST AMOC proxy the compensation for global warming and polar amplification is done by subtracting the global SST from the SG SST. By calibrating by the MOC$_z$ AMOC proxy (red curves) the optimal AMOC proxy is    SST$_{SG}$-2 SST$_{GM}$}.
\end{figure}
\end{center}

The Ceasar et al. proxy is the mean SST over the SG region subtracted the global mean in order to compensate for global warming on top of the change in the AMOC. The "translation" from the proxy SST temperature and AMOC flow is 0.26 SV/K (Caesar et al. (2018), Fig 3). Here we have taken into account that the warming is not globally homogeneous: The warming in the SG region is larger than the global mean due to polar amplification. The way we have estimated this effect is by comparing the proxy with the AMOC estimates covering the period 1957-2004 from the so-called MOC$_z$ reported in the review by Frajka-Williams et al. (2019). This shows a drop of 3 SV in that period. Minimizing the difference between the proxy SST$_{SG}$-A SST$_{GM}$  and this more direct measurement with respect to A we get A $= 1.95 \approx 2$ rather than A$=1$ used by Caesar et al. The orginal and our calibrated proxies are shown in Fig. \ref{fig:calibrated}.

\end{document}